\documentclass[aps,pre,nofootinbib,twocolumn,superscriptaddress,amsmath,amssymb]{revtex4}

\usepackage{amsmath} \usepackage{epsfig}

\def\bc{\begin{center}}
\def\ec{\end{center}}
\def\be{\begin{equation}}
\def\ee{\end{equation}}

\def\bear{\begin{eqnarray}}
\def\eear{\end{eqnarray}}

\begin{document}

\title{Space-time Thermodynamics of the Glass Transition}

\author{Mauro Merolle}

\affiliation{Department of Chemistry, University of California,
Berkeley, CA 94720-1460}

\author{Juan P. Garrahan}

\affiliation{School of Physics and Astronomy, University of
Nottingham, Nottingham, NG7 2RD, UK}

\author{David Chandler}

\affiliation{Department of Chemistry, University of California,
Berkeley, CA 94720-1460}

\begin{abstract}
We consider the probability distribution for fluctuations in dynamical
action and similar quantities related to dynamic heterogeneity.  We
argue that the so-called ``glass transition'' is a manifestation of
low action tails in these distributions where the entropy of
trajectory space is sub-extensive in time.  These low action tails are
a consequence of dynamic heterogeneity and an indication of phase
coexistence in trajectory space.  The glass transition, where the
system falls out of equilibrium, is then an order-disorder phenomenon
in space-time occurring at a temperature $T_\mathrm{g}$ which is a
weak function of measurement time.  We illustrate our perspective
ideas with facilitated lattice models, and note how these ideas apply
more generally.
\end{abstract}

\maketitle

\noindent
A glass transition, where a super-cooled fluid falls out of
equilibrium, is irreversible and a consequence of experimental
protocols, such as the time scale over which the system is prepared,
and the time scale over which its properties are observed (for reviews
see~\cite{reviews1,reviews2,reviews3}).  It is thus not a transition
in a traditional thermodynamic sense.  Nevertheless, the phenomenon is
relatively precipitous, and the thermodynamic conditions at which it
occurs depend only weakly on preparation and measurement times.  In
this paper, we offer an explanation of this behavior in terms of a
thermodynamics of trajectory space.

Our considerations seem relatively general as they are a direct
consequence of dynamic
heterogeneity~\cite{dh-reviews1,dh-reviews2,dh-reviews3} in glass
forming materials.  Our primary point is that the order-disorder of
glassy dynamics is revealed by focusing on a probability distribution
for an extensive variable manifesting dynamic heterogeneity.  There
are various choices for this variable, depending upon the specific
system under investigation.  We have more to say about this later, but
to be explicit about what can be learned, most of this paper focuses
on kinetically constrained models~\cite{Ritort-Sollich} and the
dynamical actions of these models.

We show here that due to the emergence of spatial correlations in the
dynamics, i.e. dynamic heterogeneity, action distributions have larger
low action tails than what would be expected in a homogeneous system.
These tails indicate a coexistence between space-time regions where
motion is plentiful and regions where motion is rare.  In the latter
the entropy of trajectories is subextensive in time, as previously
suggested \cite{Garrahan-Chandler}.  The glass transition, where the
system falls out of equilibrium at long but finite observation time,
is thus a disorder-order transition in space-time.  This ordering in
trajectory space is not a consequence of any underlying static
transition (for thermodynamic approaches see
\cite{Thermo1,Thermo2,Thermo3,Thermo4}).  Further, it follows from our
results here that the exponential tails observed in the aging of soft
materials, so-called intermittency
\cite{Intermittency1,Intermittency2,Intermittency3} (see also
\cite{Intermittency4,Intermittency5,Intermittency6}), are a
consequence of dynamic heterogeneity and should be seen in mesoscopic
measurements in the equilibrium dynamics of glass formers.

\bigskip

\noindent
{\bf Models and distributions of dynamical activity.} We consider two
facilitated models of glass formers to illustrate our ideas: the
lattice model of Fredrickson and Andersen
(FA)~\cite{Fredrickson-Andersen} in spatial dimension $d=1$, and its
dynamically asymmetric variant, the East model~\cite{Jackle-Eisinger}.
The FA and East models serve as a caricatures of strong and fragile
glass formers, respectively~\cite{goodies1,goodies2} both cases, there
is an energy function $J\sum_{i}n_{i}$, where $J>0$ sets the
equilibrium temperature scale, $n_{i}$ is either 1 or 0, indicating
whether lattice site $i$ is excited or not, and the sum over $i$
extends over lattice sites.  The system moves stochastically from one
micro-state to another through a sequence of single cell moves.  In
the FA model, the state of cell $i$ at time slice $t+1$, $n_{i,t+1}$,
can differ from that at time slice $t$, $n_{i,t}$, only if at least
one of two nearest neighbors, $i \pm 1,$ is excited at time $t$.  In
the East model the condition is that $n_{i+1,\,t}$ must be excited.
These dynamic constraints affect the metric of motion, confining the
space-time volume available for trajectories~\cite{Whitelam-Garrahan}.
This mimics the effects of complicated intermolecular potentials in a
dense nearly jammed material.  Excitations in this picture are regions
of space-time where molecules are unjammed and exhibit mobility.  As
such, we refer to $n_{i,t}$ as the mobility field.  For both models,
the dynamics is time-reversal symmetric and obeys detailed balance.
The equilibrium concentration of excitations, $c \equiv \langle n
\rangle = 1/(1+e^{J/T})$, is the relevant control parameter.  The
average distance between excitations sets the characteristic
lengthscale for relaxation, $\ell \approx c^{-1}$, and thus the
typical relaxation time, $\tau \approx c^{-\Delta}$, where $\Delta =
3$ for the FA model and $\Delta \approx \ln{c}/\ln{2}$ for the East
model.  See Ref.~\cite{Ritort-Sollich} for details.

\begin{figure*}
\begin{centering}
\includegraphics[width=8.0cm]{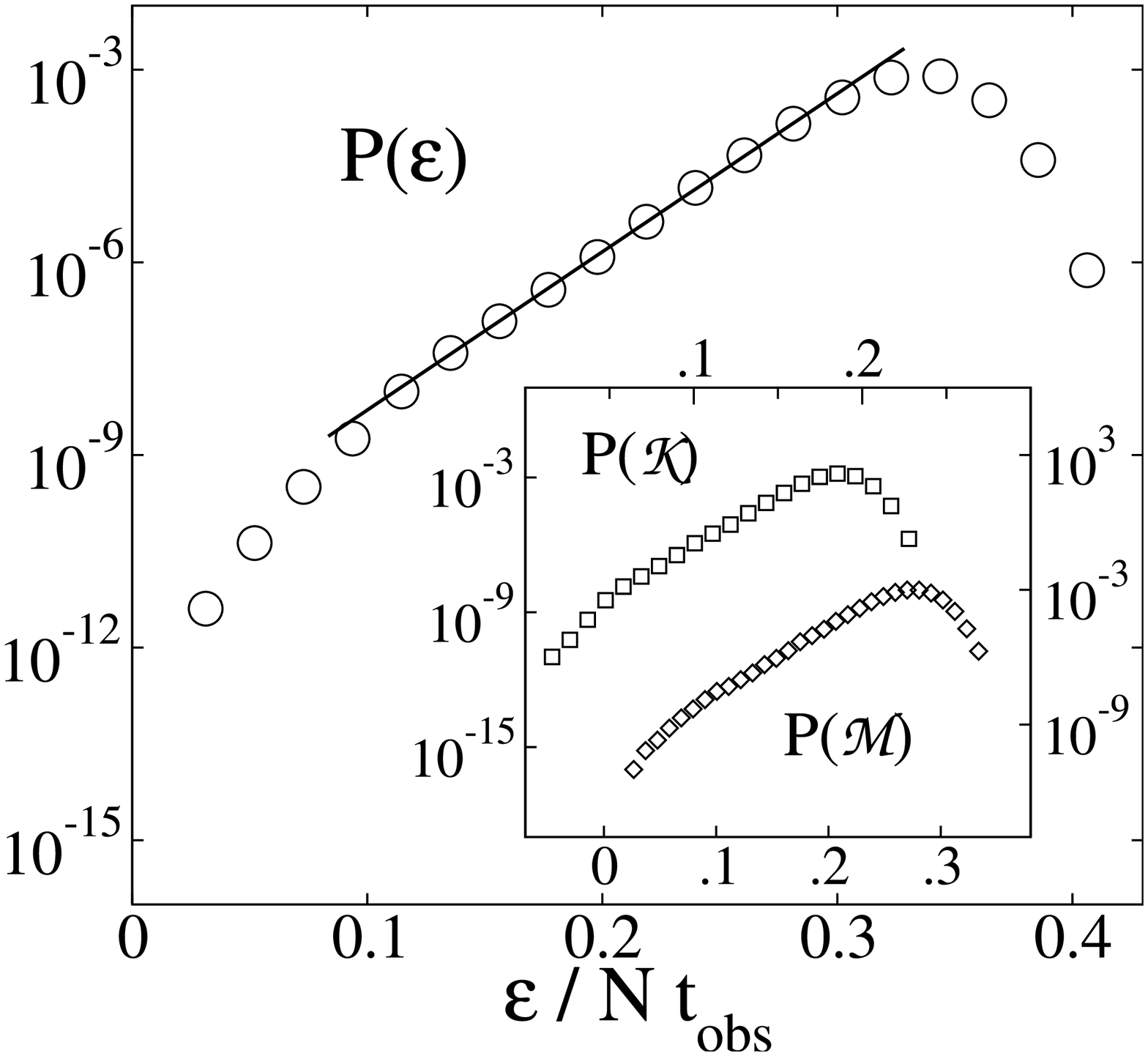}
\includegraphics[width=8.0cm]{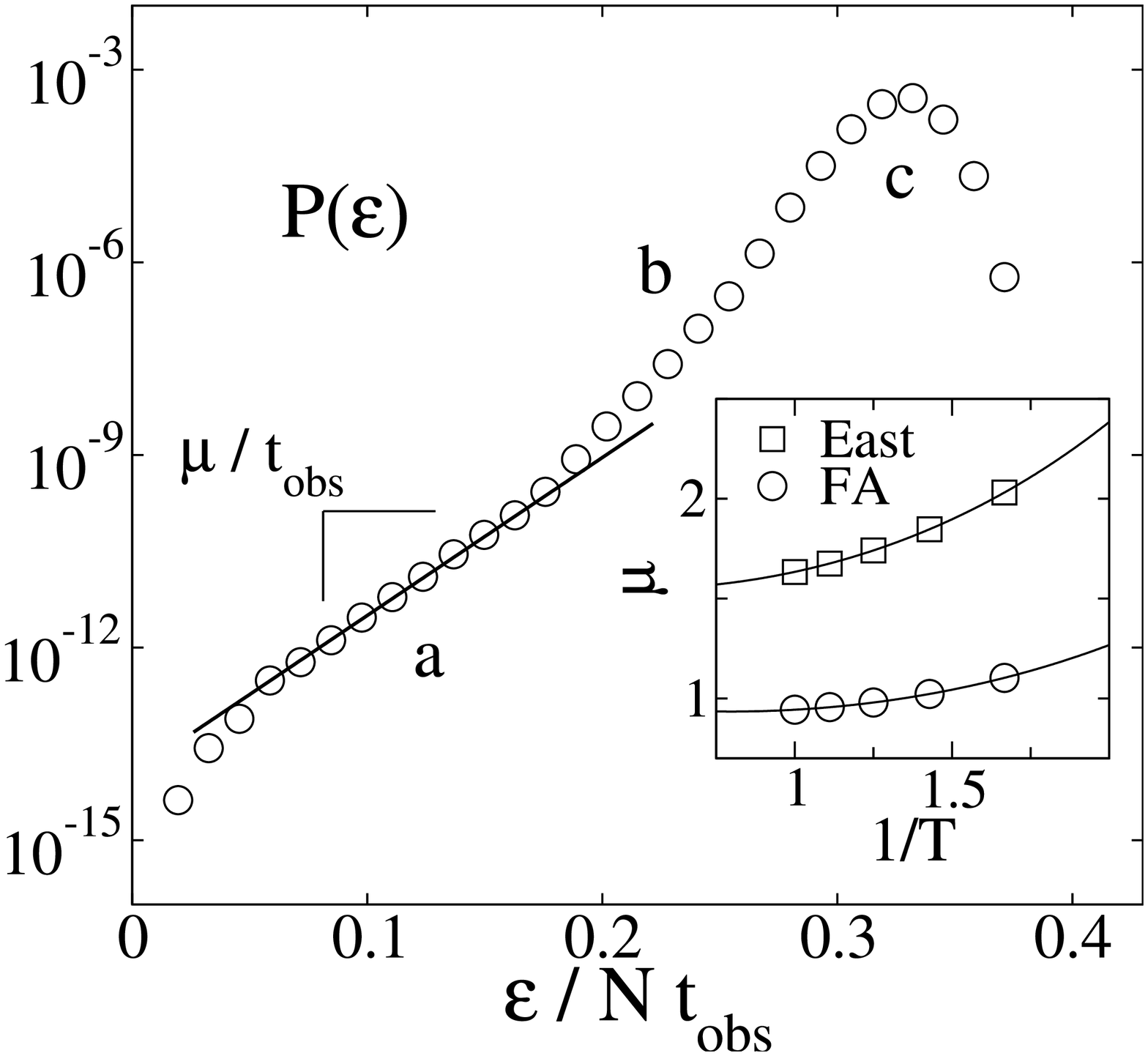}
\caption{LEFT: Probability distribution of the action $P(\mathcal{E})$
in the FA model at $T=1$, subsystem size $N=60 \approx 16~\ell$ and
$t_\mathrm{obs}=320 \approx 3 \tau$.  Inset shows probability
distribution of number of kinks, $P(\mathcal{K})$ (left and top axes),
and of number of excitations, $P(\mathcal{K})$ (bottom and right
axes).  RIGHT: $P(\mathcal{E})$ but now for $t_\mathrm{obs}=1280
\approx 12 \tau$.  The straight line indicates the exponential tail
$\exp(\mathcal{E} \mu/t_\mathrm{obs})$ (regime $a$, see text).  Inset
shows coefficient $\mu$ at different $T$ from simulation and theory
for both the FA and East models.  Statistical uncertainties from
simulations are smaller than the symbols. }
\end{centering}
\end{figure*}

\begin{figure}
\begin{centering}
\includegraphics[width=8.0cm]{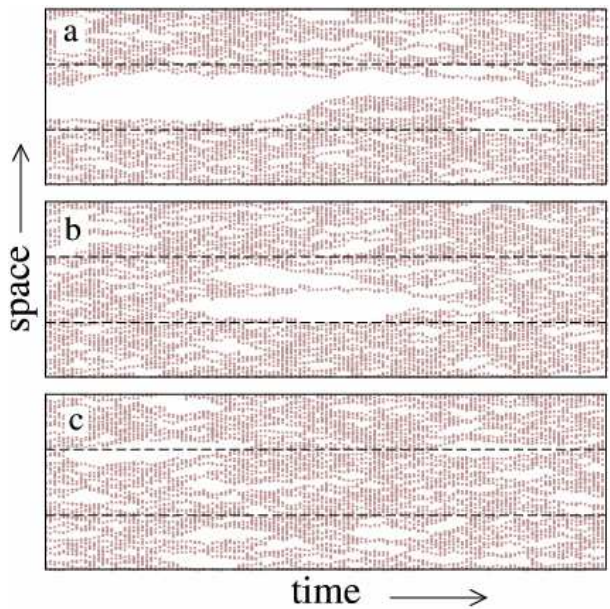}
\caption{Trajectories in the FA model corresponding to the regimes
(a), (b) and (c) of Fig.\ 1.  Unexcited and excited sites are colored
white and dark, respectively.  Space runs along the vertical
direction, and time along the horizontal direction.  Dashed lines
delimit the measured space-time volume, $N \times t_\mathrm{obs}$.  }
\end{centering}
\end{figure}

The total system under consideration has $N_{{\rm tot}}$ sites and
evolves for $t_{{\rm tot}}$ time steps, so that the total space-time
volume is $N_{{\rm tot}} \times t_{{\rm tot}}$.  Within it, we
consider a subsystem with spatial volume $N$, and observed time
duration $t_\mathrm{obs},$ and use $x(t_\mathrm{obs})$ to denote a
trajectory in that space-time volume.  For the FA and East models,
this trajectory specifies the mobility field $n_{i,t}$ for $1\leq
i\leq N$ and $0\leq t\leq t_\mathrm{obs}$.  The probability density
functional for $x(t_\mathrm{obs})$, denoted by $P[x(t_\mathrm{obs})]$,
defines an action, ${\cal E[} x(t_\mathrm{obs})]$; namely,
$P[x(t_\mathrm{obs})]\equiv \exp \left( -{\cal E}[x(t_\mathrm{obs}
)]\right)$.  In the FA and East models, ${\cal
E}[x(t_\mathrm{obs})\mathcal{]}$ is a sum over $i$ and $t$ of terms
coupling $n_{i,t}$ to mobility fields at nearby space-time
points\footnote{The microscopic action of either the FA or East models
is $\mathcal{E}[x(t)] = \mathcal{E}(x_t | x_{t-\delta t}) + \cdots +
\mathcal{E}(x_{\delta t} | x_0)-\ln[\rho(x_0)]$.  Here, $x_t \equiv \{
n_{i t} \}$ denotes a configuration at time $t$, $\rho(x_0)$ is the
distribution of initial conditions and $\mathcal{E}(x_{t+\delta t} |
x_t) \equiv - \ln{p(x_{t+\delta t} | x_t)}$, with $p(x_{t+\delta t} |
x_t)$ the probability for an elementary microscopic move in time
$\delta t$.  In a Monte Carlo simulation, for example, $\delta t=1/N$
and the move is an attempted spin-flip.  The explicit form of the
operator $\mathcal{E}(x_{t+\delta t} | x_t)$ is given in
\cite{Garrahan-Chandler}.  Here, we use continuous time MC
\cite{Bortz} where all attempted moves are accepted: site $i$ is
chosen to flip at time $t$, $n_i \to 1-n_i$, with probability $(\delta
t) [n_i + (1-n_i) e^{-1/T}]$, and time is increased by $\delta t =
(N_{\rm 1f} + N_{\rm 0f} e^{-1/T})^{-1}$, where $N_{\rm 1f}$ and
$N_{\rm 0f}$ are the number of facilitated up and down sites at time
$t$, respectively.  The corresponding contribution to the action is
$\mathcal{E}(x_{t+\delta t} | x_t) = - \ln{\delta t} + T^{-1} (1-n_i)
- \ln N$ (the last term is added to remove trivial $\ln N$
dependences, also present in standard MC).}.  As such, its average
value will be extensive in $N\times t_\mathrm{obs}$. This extensivity
is a general property for any system with Markovian dynamics and
short-ranged interparticle forces.

The distribution function for the action is
\begin{equation}
P(\mathcal{E}) = \left\langle \delta \left(
\mathcal{E-E[}x(t_\mathrm{obs})]\right) \right\rangle = \Omega \left(
\mathcal{E}\right) \exp \left( -\mathcal{E}\right) , \label{P(E)}
\end{equation}
where the pointed brackets indicate average over the ensemble of
trajectories of length $t_\mathrm{obs}$, and $\Omega \left(
\mathcal{E}\right) =\Omega \left( \mathcal{E};N,t_\mathrm{obs}\right)
$ is the number of such trajectories with action $\mathcal{E}$.  When
$N$ is much larger than any dynamically correlated volume in space,
and $t_\mathrm{obs}$ is much larger than any correlated period of
time,
\begin{equation}
\ln{\Omega\left( \mathcal{E},N,t_\mathrm{obs}\right)} \equiv s \left(
\mathcal{E};N,t_\mathrm{obs}\right) N t_\mathrm{obs} ,
\end{equation}
will be extensive in the space-time volume $Nt_\mathrm{obs}$.  In this
case, the entropy per space-time point, $s \left(
\mathcal{E};N,t_\mathrm{obs}\right)$, will be intensive.  Similarly,
and for the same reasons, the mean square fluctuation,
$\chi(N,t_\mathrm{obs}) = \langle \mathcal{E}^2 \rangle - \langle
\mathcal{E} \rangle^2$, is extensive for large enough
$Nt_\mathrm{obs}$. The onset of this extensivity with respect to
$t_\mathrm{obs}$ can be viewed as an order-disorder phenomenon, as we
will discuss shortly.

The distribution function for the action can be obtained from
simulation by running trajectories and creating a histogram for the
logarithm of the probabilities for taking steps in the trajectory.
Far into the wings of the distribution, satisfactory statistics is
obtained with transition path sampling~\cite{TPS}. This methodology
allows us to carry out umbrella sampling~\cite{Frenkel-Smidt} applied
to trajectory space\footnote{A trajectory of the total system is
accepted if the action of the subsystem computed from that trajectory
lies between ${\cal E} _{1}$ and ${\cal E}_{2}$. A new trajectory is
created from the accepted trajectory by implementing shooting and
shifting moves~\cite{TPS}. The new trajectory is accepted if the
action of the subsystem again lies between ${\cal E}_{1}$ and ${\cal
E}_{2}$, and rejected otherwise. The ensemble of trajectories created
by repeating this process many times provides the distribution
function for the action for the subsystem when that action is between
${\cal E}_{1}$ and ${\cal E}_{2}$. Yet another part of the
distribution is created by performing the sampling again, this time
for subsystem actions in a different range, say between ${\cal E}_{3}$
and ${\cal E}_{4}$. Eventually, the distribution is mapped out over
the entire range of interest by splicing together many of these
partial distributions.}.  Figure 1 illustrates the
$P\left( \mathcal{E}\right)$'s we have obtained in this way for the FA
and East models.  While the bulk of the distributions is Gaussian, for
values of the action sufficiently smaller than $\langle \mathcal{E}
\rangle$ they display exponential tails.  Similar distributions are
found for other measures of dynamic activity, such as the number of
transitions or kinks in a space-time volume,
$\mathcal{K}=\sum_{i,t}\left[ n_{i,t}\,\left( 1-n_{i,t+\delta
t}\right) +\left( 1-n_{i,t}\right) \,n_{i,t+\delta t}\right]$, or the
total number of excitations $\mathcal{M}=\sum_{i,t} n_{i,t}$.  This is
shown in the inset to Fig.\ 1(left).

\begin{figure*}
\begin{centering}
\includegraphics[width=8.0cm]{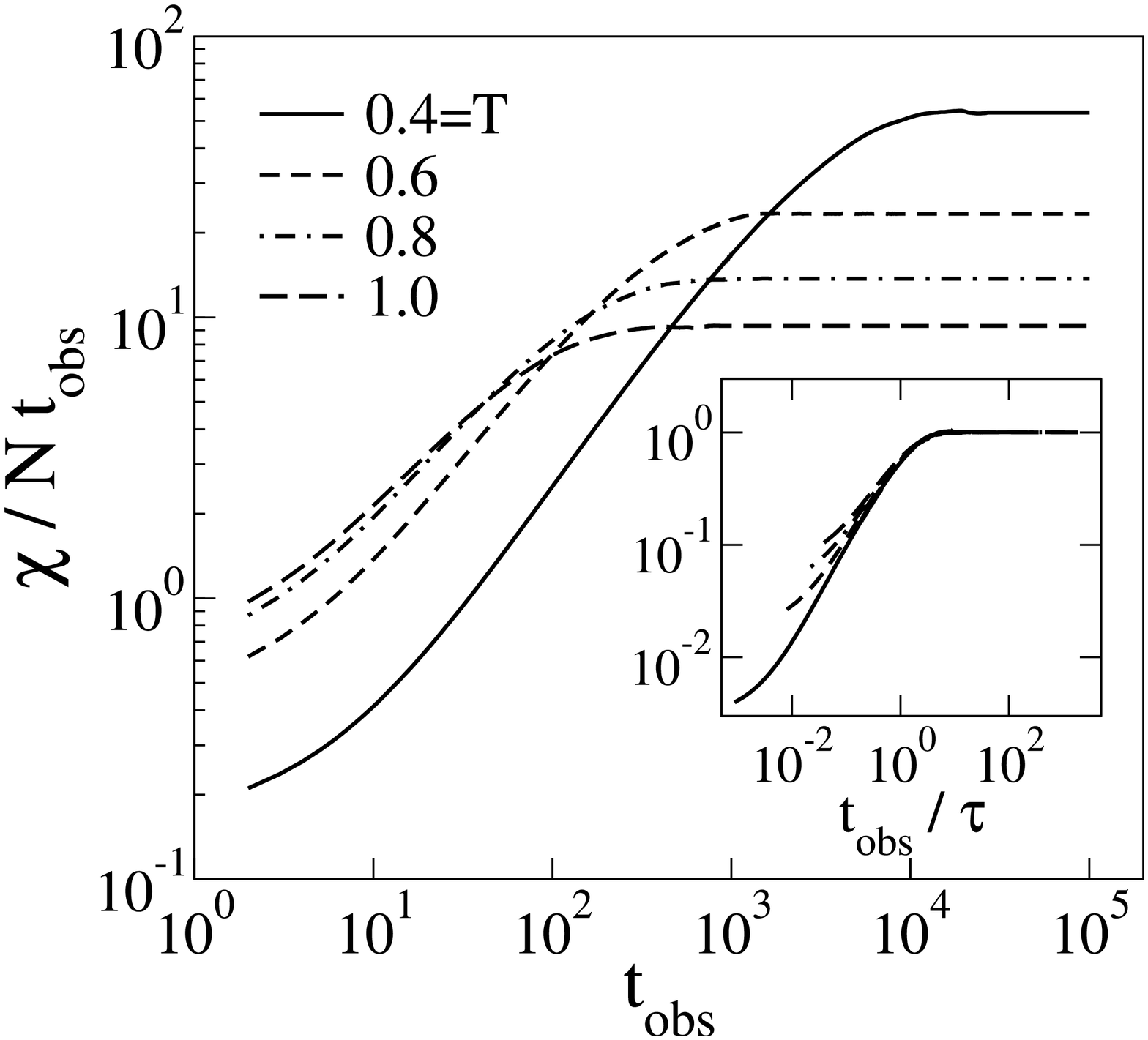}
\includegraphics[width=8.0cm]{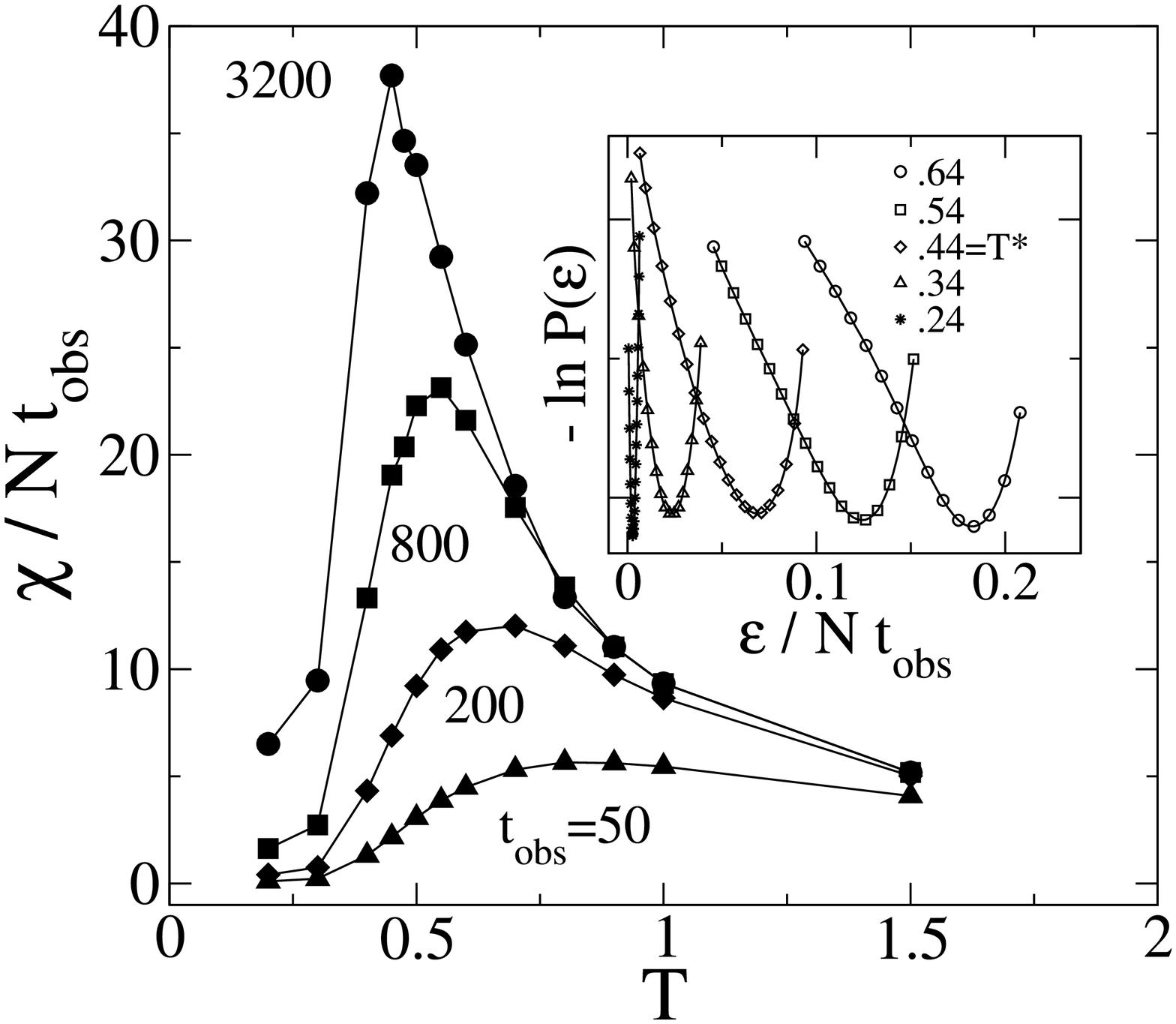}
\caption{LEFT: Action susceptibility per unit length and time, $\chi /
N t_\mathrm{obs}$, as a function of observation time $t_\mathrm{obs}$
for constant temperature $T$, in the FA model.  Inset shows scaled
susceptibility, $(t_\mathrm{obs}^{-1} \chi) / [\lim_{t_\mathrm{obs}
\to \infty} (t_\mathrm{obs}^{-1} \chi)]$ as a function of
$t_\mathrm{obs}/\tau$.  RIGHT: Action susceptibility per unit length
and time, $\chi / N t_\mathrm{obs}$, but now as a function of $T$ at
constant $t_\mathrm{obs}$.  Inset shows free energies of trajectories,
$-\ln{P(\mathcal{E})}$, for observation time $t_\mathrm{obs}=3200$ at
different temperatures.  System sizes are $N=16 c^{-1}$.  For
$t_\mathrm{obs}=3200$, $T^* \approx 0.44$.  Statistical uncertainties
from simulations are smaller than the symbols.}
\end{centering}
\end{figure*}

\begin{figure}
\begin{centering}
\includegraphics[width=8.0cm]{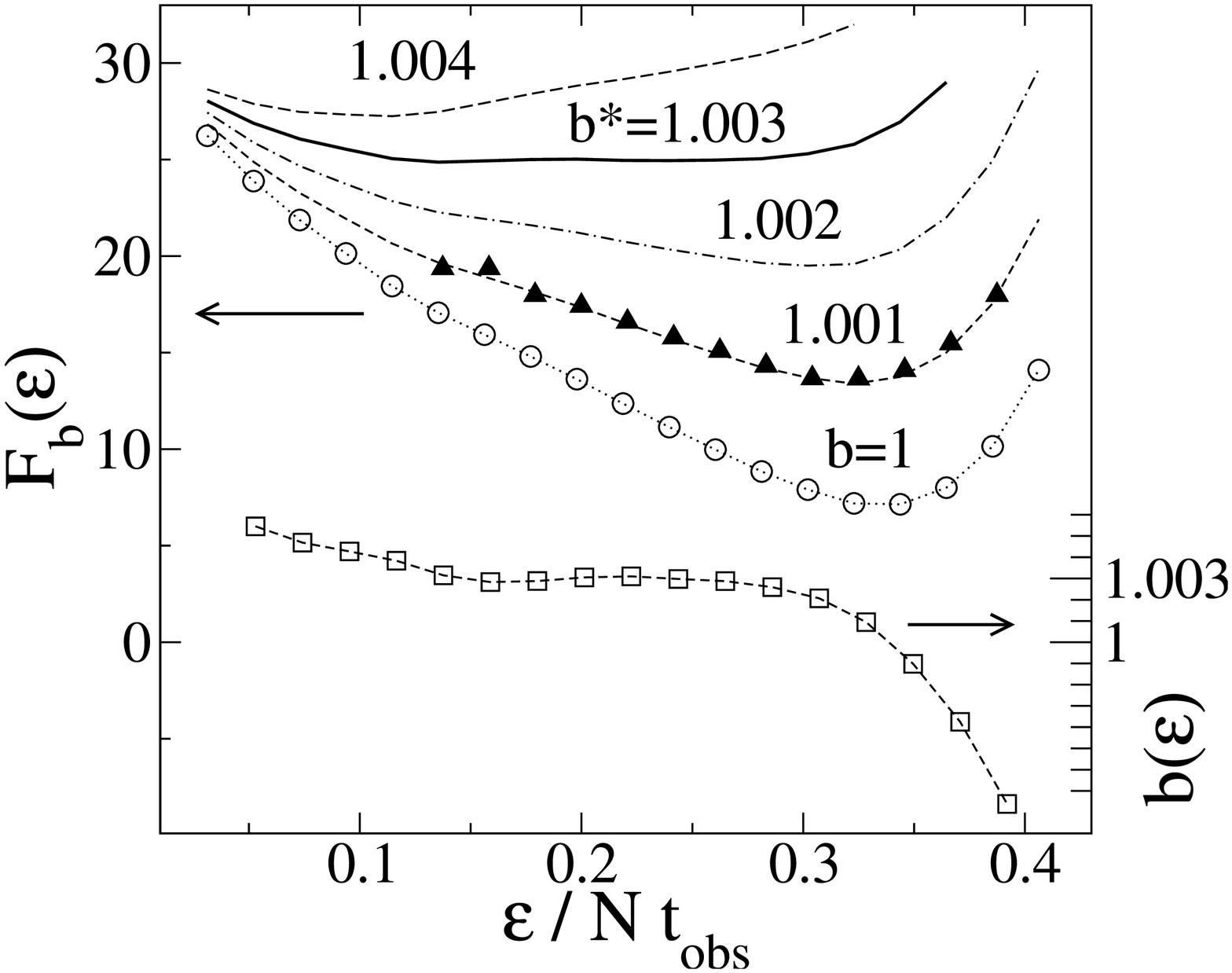}
\caption{Free energies of trajectories, $F_b(\mathcal{E})$, for
different values of the parameter $b$.  The curve for $b=1$ is that of
Fig.\ 1(left), with $T=1$, $N=16 c^{-1}$, $t_\mathrm{obs}=320$.  The
curves with $b=1.001,~1.002,~1.003$ are obtained by adding $(b-1)$ to
the $b=1$ curve.  Triangles show the free energy computed by
simulation at $b=1.001$ using transition path sampling combined with
free energy perturbation theory.}
\end{centering}
\end{figure}

\bigskip

\noindent
{\bf Dynamic heterogeneity and susceptibility.} Figure 2 makes vivid
the fact that the exponential tails in these distributions are
manifestations of dynamic heterogeneity.  The extended bubble or
stripe in trajectory (a) illustrates a volume of space-time that is
absent of excitations.  Because it extends throughout the pictured
time frame, the statistical weight for this excitation void is
dominated by the probability to observe this void at the initial time
$t=0$.  As excitations at a given time slice are uncorrelated, this
probability is Poissonian, $\exp(-c L )$, where $L$ is the width of
the stripe.  In the presence of a bubble occupying a space-time volume
$L~t_\mathrm{obs}$, the net action is $\mathcal{E} \approx
t_\mathrm{obs}~(N-L)~\varepsilon$, where $\varepsilon$ is the average
action per unit space-time.  In this regime, the probability density
for the action is then $\exp \left[ -c L\left( \mathcal{E}\right)
\right] \propto \exp \left( c~\mathcal{E}/t_\mathrm{obs}\varepsilon
\right)$.  This proportionality explains why the slope of the
exponential tail scales linearly with inverse time. The specific value
of that slope is determined from evaluating the action
density. Performing the requisite average of the action gives
$\varepsilon \equiv \langle \mathcal{E} \rangle / N t_\mathrm{obs}
\approx -f(c) \, c \, \ln{c}$, for small $c$, where $f(c) \approx 2c$
for the FA model, and $f(c) \approx c$ for the East model.  Hence, to
the degree that this picture is correct, the slope $\mu
/t_\mathrm{obs}$ in Fig.~1 should coincide with $\mu
=c/\varepsilon$. The inset to Fig.~1(right) shows that this
relationship holds to a good approximation.  Thus, the exponential
tail in $P\left(\mathcal{E}\right)$ manifests structures in space-time
that depend upon initial conditions over a time frame
$t_\mathrm{obs}$.  It also follows that the entropy for trajectories
with action that falls in the exponential tails is non-extensive in
time $t_\mathrm{obs}$.

The tails in $P(\mathcal{E})$ are statistically negligible when
$t_\mathrm{obs}$ is large compared to the relaxation time $\tau$, but
they are dominant when $t_\mathrm{obs} \lesssim \tau$.  The latter
regime is where the mean square fluctuation or susceptibility,
$\chi(N,t_\mathrm{obs})$, is non-linear in time.  Figure 3(left) shows
the growth of this quantity with respect to $t_\mathrm{obs}$.  The
susceptibility per unit space-time increases with increasing
$t_\mathrm{obs}$ because increasing time allows for increased
fluctuations.  With the onset of decorrelation, when $t_\mathrm{obs} >
\tau$, the susceptibility becomes extensive, and
$\chi(N,t_\mathrm{obs}) / Nt_\mathrm{obs}$ becomes a constant.  This
plateau or constant value increases with decreasing $T$ because
fluctuations become more prevalent as $c$ decreases. Indeed, the FA
and East models approach dynamic criticality as $c \to
0$~\cite{Whitelam-et-al}.

While the plateau value of $\chi \left( N,t_\mathrm{obs} \right) /
Nt_\mathrm{obs}$ increases with decreasing $T$ or $c$, the time for
the onset to extensivity also increases.  As a result, if
$t_\mathrm{obs}$ is kept fixed and $T$ is varied, $\chi \left(
N,t_\mathrm{obs} \right)$ will have a maximum.  This behavior is
illustrated in Fig.\ 3(right).  The extremum will approach a
singularity in the limit of criticality, $t_\mathrm{obs}\rightarrow
\infty ,$ $ c\rightarrow 0$.  For $t_\mathrm{obs}$ fixed and finite,
the extremum is located at a finite temperature, $T^{*}$, namely,
$\tau \left( T^{*}\right) \sim t_\mathrm{obs}$, where $\tau \left(
T\right)$ denotes the equilibrium relaxation time at the temperature
$T$.  Accordingly, $T^{*}$ is a glass transition temperature.  In
particular, below this temperature, the system cannot equilibrate on
time scales as short as $t_\mathrm{obs}$.  In other words, with
observation times no longer than this, the system will have fallen out
of equilibrium. For the Arrhenius FA model, $\ln{\tau(T)} \varpropto
1/T$, so that $ T^{*}$ in this case varies logarithmically with
observation time $t_\mathrm{obs}$.  For the super-Arrhenius East
model, $\ln{\tau(T)} \propto 1/T^{2} $, $T^{*}$ is even more weakly
dependent upon observation time, going as
$1/\sqrt{\ln{t_\mathrm{obs}}}$.  It is due to this weak dependence on
observation time that the glass transition is very nearly a material
property, being confined to a narrow range of temperatures.

Similar behaviors are expected for distributions of analogous
quantities in continuous force systems.  One such quantity is
$\mathcal{Q}=\sum_{j\in V}\int_{0}^{t_\mathrm{obs}}dt\exp \left\{
ik\cdot \left[ r_{j}\left(t+\delta t\right) -r_{j}\left( t\right)
\right] \right\}$, where $r_{j}\left(t\right)$ refers to the position
of the $j$th particle at time $t$, and the sum includes all those
particles $j$ that are in volume $V$ at time $0$.  The wave-vector $k$
and time lag $\delta t$, while microscopic, should be such that
variation in $\mathcal{Q}$ is due to diffusion of particles, not
simply vibrational motion. In the past, studies of
dynamic heterogeneity have considered the susceptibility for
$d\mathcal{Q}/ d t_\mathrm{obs}$, regarding the time lag $\delta t$ as
a variable that can be large
\cite{Whitelam-et-al,Toninelli-et-al}.  The effect of this
variability and the differentiation is to focus on an object that is
not extensive in $t_\mathrm{obs}$, thus obscuring order-disorder in
space-time.

Not all variables extensive in space and time are
useful.  This situation is familiar in standard phase transition
theory.  Phase transitions are recognized only through physically
motivated choices of variables that are then shown to reflect the
transition.  For example, consider Langevin dynamics,
$\dot{\vec{x}}(t) = \vec{f}[\vec{x}(t)] + \vec{\eta}(t)$, where
$\vec{\eta(t)}$ is a Gaussian random force.  In this case, the
contribution to the action for a transition in infinitesimal time $dt
\to 0$ is, up to a constant, $- \ln p[\vec{x}(t+dt) | \vec{x}(t)] = dt
\{ \dot{\vec{x}}(t) - \vec{f}[\vec{x}(t)] \}^2 / 4T$, which is
distributed as $\frac{dt}{4T} \vec{\eta}^2$ independently of the
forces $\vec{f}$.  The net action for a trajectory is simply the
accumulation of these infinitesimal steps, and will be therefore
distributed in a trivial manner.  It is insensitive to the phenomenon
we are after.  The action for non-linear functions or functionals of
the Langevin $\vec{x}(t)$, however, need not be trivial~\cite{Langevin}.  For
variables that manifest dynamic heterogeneity, in particular, a
meaningful observable can be constructed out of the probability for
finite (not infinitesimal) steps: $P[\vec{x}(t+\delta t) | \vec{x}(t)]
= \int dx_{t+dt} dx_{t+2dt} \cdots p[\vec{x}(t+dt) | \vec{x}(t)]
\cdots p[\vec{x}(t+\delta t) | \vec{x}(t+\delta t-dt)]$, where $\delta
t$ is an appropriate coarse-graining time.  The integrals over
intermediate states make the action for finite moves,
$\mathcal{E}[\vec{x}(t+\delta t) | \vec{x}(t)] \equiv - \ln
P[\vec{x}(t+\delta t) | \vec{x}(t)]$, dependent on the actual forces
$\vec{f}$, and the distribution for this action will be system
dependent.  The case of deterministic dynamics is similar in that the
microscopic action is always vanishing, and a coarse-grained action is
needed to reveal interesting behaviour.  Whether or not this
coarse-graining is practical, other variables, such as $\mathcal{Q}$,
are clearly useful and practical.

\bigskip

\noindent
{\bf Order-disorder in space-time.}  Above $T^{*}$, the system is
equilibrated and disordered with respect to order parameters like
$\mathcal{E}$ or $\mathcal{K}$ or $\mathcal{Q}$.  Below $T^{*}$, the
system is ordered with respect to these quantities, and the order
manifests dependence upon initial conditions.  In view of the
anomalous behavior of $\chi \left(N,t_\mathrm{obs};T\right)$ near
$T=T^{*}$, this dependence can be manipulated in a fashion familiar in
the context of equilibrium phase transitions.  To this end, it is
useful to define $b\equiv \partial \ln \Omega \,/\partial \mathcal{E}$
and a corresponding free energy $F_{b}(\mathcal{E}) \equiv
-\ln\Omega\left( \mathcal{E}\right)+b\mathcal{E}= -\ln P\left(
\mathcal{E}\right)+(b-1)\mathcal{E} $.  The parameter $b$ is to
trajectory space what reciprocal temperature is to state space.  In
particular, an extremum in $\chi \left( T\right) $ corresponds with a
rapid change in $\left\langle \mathcal{E}\right\rangle $ with respect
to $T$, which in turn implies a bistable free energy, $F_{b}\left(
\mathcal{E}\right) $, near $b=1$ and $T=T^{*}$.

Fig.\ 4 illustrates this behavior.  Coexistence takes place at $b= b^*
> 1$, the precise value depending upon $T$ and $t_\mathrm{obs}$.  The
free energy for $b>1$ can be computed from path sampling with a form
of thermodynamic perturbation theory~\cite{Frenkel-Smidt}, running
trajectories as before (i.e., with $b=1$) and averaging the functional
$\exp \left( -\Delta b{\cal E}[x(t_\mathrm{obs} )]\right)$.  In the
case illustrated, $b^* = 1.003$, very close to the physical value
$b=1$.  The basin at low action coincides with the exponential tail
and thus correlation with initial conditions throughout the observed
time frame.

For an equilibrium system, $b=1$.  In this case, there will be a true
dynamical singularity for $t_\mathrm{obs} \to \infty$ only at $T=0$.
In particular, from the exponential tails of $P(\mathcal{E})$, we know
that $b^* = 1+\mu / t_\mathrm{obs}$, so that $b^* \to 1$ when $c \to
0$ with $t_\mathrm{obs} / \tau \to \mathrm{finite}$.  An interesting
question is whether it is possible to define a dynamical protocol,
perhaps through some form of external driving, which would allow one
to tune $b$ from $b=1$ to $b=b^*$, and thus observe this dynamical
transition at $T>0$.

\bigskip

We thank Hans Andersen and Jorge Kurchan for helpful comments on
 earlier drafts of this paper.  This work was supported by the NSF, by
 DOE grant no.\ DE-FE-FG03-87ER13793, by EPSRC grants no.\
 GR/R83712/01 and GR/S54074/01, and University of Nottingham grant
 no.\ FEF 3024.

\end{document}